\documentclass[prd,twocolumn,superscriptaddress,showpacs,preprintnumbers,amsmath,amssymb,nofootinbib]{revtex4-1}
\usepackage[dvipdfmx]{graphicx}
\usepackage{dcolumn}
\usepackage{bm}
\usepackage[dvipdfmx]{color}

\usepackage{color}

\begin{document}


\title{Three-quark potential and Abelian dominance of confinement in SU(3) QCD}
\author{Naoyuki~Sakumichi}
\affiliation{Theoretical Research Division, Nishina Center, RIKEN, Wako, Saitama 351-0198, Japan}
\author{Hideo~Suganuma}
\affiliation{Department of Physics, Kyoto University, 
Kitashirakawaoiwake, Sakyo, Kyoto 606-8502, Japan}
\date{\today}
\begin{abstract}
We study the baryonic three-quark (3Q) potential 
and its Abelian projection in terms of 
the dual-superconductor picture in SU(3) quenched lattice QCD.
The non-Abelian SU(3) gauge theory is projected onto 
Abelian U(1)$^2$ gauge theory in the maximal Abelian gauge.
We investigate the 3Q potential and its Abelian part 
for more than 300 different patterns of static 3Q systems in total at $\beta=5.8$ on $16^332$ and at $\beta=6.0$ on $20^332$ with $1000$--$2000$ gauge configurations.
For all the distances, both the 3Q potential and Abelian part 
are found to be well described by the Y ansatz, i.e., 
two-body Coulomb term plus three-body Y-type linear term 
$\sigma_{3\mathrm{Q}} L_{\mathrm{min}}$, where $L_{\mathrm{min}}$ 
is the minimum flux-tube length connecting the three quarks.
We find equivalence between the three-body string tension 
$\sigma_{3\mathrm{Q}}$ and its Abelian part $\sigma_{3\mathrm{Q}}^{\rm Abel}$ 
with an accuracy within a few percent deviation, i.e., $\sigma_{3\mathrm{Q}} \simeq \sigma_{3\mathrm{Q}}^{\rm Abel}$, which means Abelian dominance of the quark-confining force in 3Q systems.
%
%
\end{abstract}

\maketitle



\section{Introduction}
%
Quark confinement has been one of the most important long-standing issues remaining in theoretical physics \cite{QCD}
since the concept of quarks was introduced in the 1960s.
In fact, quarks cannot be observed individually and are confined in color-singlet combinations of mesons or baryons.
In particular, the nucleon, the lightest baryon, is one of the main ingredients of the matter in our real world, 
and, therefore, the quark confinement in baryons or three-quark (3Q) systems would be fairly important in modern physics, 
as well as in mesons or quark-antiquark (Q$\bar{\rm Q}$) systems.
Furthermore, the three-body force among three quarks is a ``primary'' force reflecting the SU(3) gauge symmetry in quantum chromodynamics (QCD) \cite{TSNM01, TSNM02}, 
while the three-body force appears as a residual interaction in most fields of physics. 
Nevertheless, the quark interaction in baryonic 3Q systems 
\cite{TSNM01, TSNM02, C2004} has not been investigated so much, 
in contrast with many lattice studies on Q$\bar{\mathrm{Q}}$ systems 
\cite{QCD,BSS93,B01}.

\begin{figure}[b]
\centering
\includegraphics[width=8.6cm,clip]{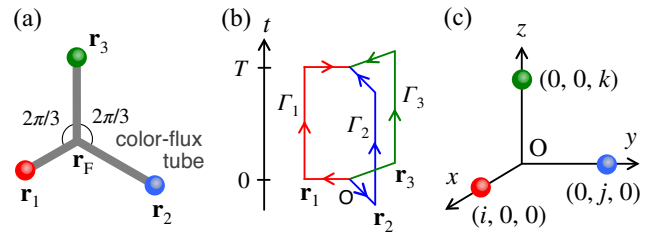}
\caption{
(a) The flux-tube configuration of the three-quark system with the minimal value of the total flux-tube length. 
There appears a physical junction linking the three flux tubes at the Fermat point $\mathbf{r}_F$.
(b) The trajectory of the 3Q Wilson loop $W_{3\mathrm{Q}}$.
The three quarks are generated at $t=0$, are spatially fixed 
in $\mathbb{R}^3$ for $0<t<T$, and are annihilated at $t=T$.
(c) The configuration of static three-quark sources 
in our lattice QCD simulations. }
\label{Fig:1}
\end{figure}

In SU(3) quenched lattice QCD,
%
%
the static Q$\bar{\rm Q}$ \cite{BSS93} 
and 3Q \cite{TSNM01, TSNM02,OST05} potentials are found to be 
well reproduced by 
\begin{gather} 
V(r) = \sigma r - \frac{A}{r} +C,  \label{eq:Cornell}  \\
%
V_{3\mathrm{Q}} ({\bf r}_1, {\bf r}_2, {\bf r}_3)
= \sigma_{3\mathrm{Q}}  L_{\rm min}-\sum_{i<j} \frac{A_{3\mathrm{Q}} }{|{\bf r}_i-{\bf r}_j|}+C_{3\mathrm{Q}},
\label{eq:Y-ansatz}
\end{gather}
respectively.
Here, ${\bf r}_1, {\bf r}_2$, and ${\bf r}_3$ are the positions of 
the three quarks, and $L_{\mathrm{min}}$ is the minimum flux-tube length 
connecting the three quarks as shown in Fig.~{\ref{Fig:1}}(a).
The form (\ref{eq:Y-ansatz}) is called the Y ansatz \cite{TSNM02}.
These functional forms (\ref{eq:Cornell}) and (\ref{eq:Y-ansatz}) indicate 
the flux-tube picture \cite{CNN79} 
on the confinement mechanism.
In fact, the lattice QCD simulations 
\cite{QCD,IBSS03, DIK04_3Q, BS04, CCB13, CCCP14} 
on the action density in the presence of a static Q$\bar{\mathrm{Q}}$ or 3Q system have actually shown the flux-tube formation; that is, 
valence quarks are linked by the color flux tube 
as a quasi-one-dimensional object.
Here, the strength of quark confinement is controlled 
by the string tension of the flux tube, $\sigma$ or $\sigma_{3{\rm Q}}$.
We also note that the baryonic 3Q system has recently received attention in the context of the holographic description of 
strong interactions, e.g., AdS/QCD effective string theories \cite{A08}.

The difficulty in deriving quark confinement directly from QCD is considered 
to originate from non-Abelian dynamics and nonperturbative features of QCD, 
which are quite different from the case of quantum electrodynamics (QED).
However, it remains unclear whether quark confinement is peculiar to 
the non-Abelian nature of QCD or not.

As an interesting idea of quark confinement, 
Nambu, 't~Hooft, and Mandelstam proposed 
an Abelian theory of the dual superconductor for the confinement mechanism 
\cite{NtHM} in the 1970s. 
In the dual-superconductor picture, 
the squeezing of the color-electric flux among quarks 
is realized by the dual Meissner effect as the result of 
condensation of color-magnetic monopoles.
(Note here that monopole condensation and its relevant role 
for confinement have been analytically pointed out by Seiberg and 
Witten in the $N=2$ supersymmetric version of 
the Yang--Mills theory \cite{SW94}.)

\begin{figure}[t]
\centering
\includegraphics[width=8.4cm,clip]{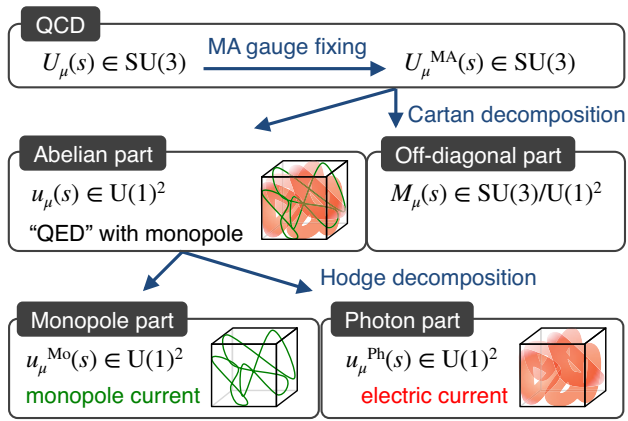}
\caption{
Schematic figure of Abelianization of QCD and the dual-superconductor scenario of confinement.
In the MA gauge, QCD becomes Abelian-like, and the monopole current topologically appears. 
By the Hodge decomposition, the QCD system can be divided into the monopole and the photon parts. 
The monopole part has confinement, chiral symmetry breaking, and instantons,
while the photon part does not have all of them \cite{SNW94,M95,W95,S95}.}
\label{Fig:2}
\end{figure}

As for the possible connection between the dual superconductor and QCD, 't~Hooft proposed a concept of ``Abelian projection'' as 
an infrared Abelianization scheme of QCD \cite{tH81,EI82}, 
where the magnetic monopole topologically appears. 
And 't~Hooft also conjectured that long-distance physics such as confinement 
could be realized only by Abelian degrees of freedom in QCD \cite{tH81}, 
which is called ``(infrared) Abelian dominance''. 
Actually, in the maximally Abelian (MA) gauge \cite{KSW87,SY90,SNW94,AS99}, 
QCD becomes Abelian-like 
as a result of a large off-diagonal gluon mass of about 1GeV \cite{AS99}, 
and the monopole current topologically appears \cite{KSW87}. 
(See Fig.\ref{Fig:2}.)
By using the Hodge decomposition, the QCD vacuum can be divided 
into the monopole and the photon parts. 
The lattice QCD studies demonstrate that 
the monopole part has confinement \cite{SNW94}, 
chiral symmetry breaking \cite{M95,W95} 
and instantons \cite{S95},
while the photon part does not have all of them.

\begin{figure}[t]
\centering
\includegraphics[width=8.0cm,clip]{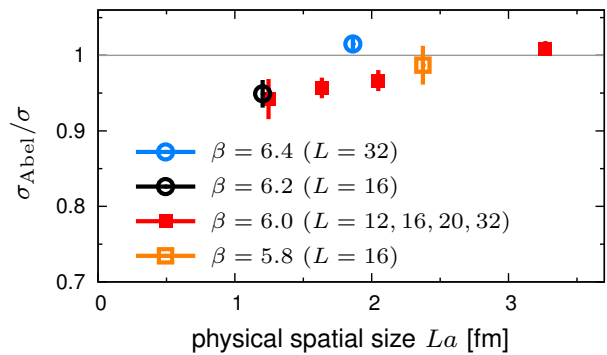}
\caption{
Physical spatial-size dependence of $\sigma_{\rm Abel}/\sigma$ taken from Ref.~\cite{SS14}.
Here, $\sigma$ and $\sigma_{\rm Abel}$ are the string tensions of the Q$\bar{\mathrm{Q}}$ potential for SU(3) QCD and the Abelian part, respectively.
Perfect Abelian dominance ($\sigma_{\rm Abel}/\sigma \simeq 1$) seems to be realized when the spatial size $La$ is larger than about $2$~fm.
In this paper, we investigate the corresponding string tensions of 3Q potentials for $\beta=5.8$ on $16^332$ and $\beta=6.0$ on $20^332$ lattices.
%
%
  %
}
 \label{Fig:PRDR}
\end{figure}

Many lattice QCD studies have remarkably shown Abelian dominance of 
the confining force in static Q$\bar{\mathrm{Q}}$ systems in the MA gauge: 
the string tension $\sigma$ is reproduced 
by the Abelian-projected one 
$\sigma^\mathrm{Abel}$ 
in both SU(2) \cite{KSW87,SY90,SNW94,AS99} 
and SU(3) \cite{STW02,DIK04_QQ} color QCD.
Recently, in the SU(3) quenched lattice QCD,
we found perfect Abelian dominance \cite{SS14} of the quark-confining force in Q$\bar{\mathrm{Q}}$ systems; i.e., 
the confining force is entirely reproduced only with the Abelian sector, $\sigma^\mathrm{Abel} \simeq \sigma$.
(To be precise, e.g., $\sigma^\mathrm{Abel} /\sigma = 1.01(1)$ and $1.00(2)$ for $\beta=6.0$ on $32^4$ lattice and $\beta=5.8$ on $16^332$ lattice, respectively.)
To obtain $\sigma^\mathrm{Abel} \simeq \sigma$, it is necessary to use 
(i) a larger numbers of gauge configurations, 
(ii) both on-axis and several types of off-axis data, 
and (iii)  large-volume lattices of more than about $2$~fm.
In particular, the use of large physical-volume lattices is essential, 
as is shown in Fig.~\ref{Fig:PRDR} \cite{SS14}.
Moreover, very recently, perfect Abelian dominance was found also 
in SU(2) quenched QCD \cite{KKS15}.
[The authors of Ref.~\cite{KKS15} reported $\sigma^\mathrm{Abel} /\sigma = 1.02(2)$ for $\beta=2.5$ on $24^4$ lattice, where the physical volume is $La\simeq 2.0$~fm.]
These observations of $\sigma^\mathrm{Abel} \simeq \sigma$
indicate that the Abelianization of QCD can be realized without loss of the quark-confining force via the MA projection.

In this paper, we investigate whether quark confinement in the baryonic 3Q potential is entirely kept in the Abelian sector of QCD in the MA gauge and find this to be true at the quenched level.
%
%
%
%
Despite the physical importance of baryons, 
there are very little studies about Abelian dominance in the baryonic 3Q potential \cite{DIK04_3Q}
because the previous lattice studies of Abelian dominance were performed mainly 
for simplified SU(2) color QCD, where the color structure of SU(2) baryons (QQ) are the same as that of mesons (Q$\bar{\mathrm{Q}}$).
In a pioneering study, 
Bornyakov~{\it et~al.} \cite{DIK04_3Q} reported 
approximate Abelian dominance of the string tension in the 3Q potential, 
$\sigma_{3\mathrm{Q}}^\mathrm{Abel} / \sigma_{3\mathrm{Q}} = 0.83(3)$, 
using the simulated annealing algorithm to avoid Gribov copy effects, 
on a $16^3 32$ lattice at $\beta=6.0$. 
However, from the results of $\sigma^\mathrm{Abel} \simeq \sigma$ 
in mesonic Q$\bar{\mathrm{Q}}$ cases \cite{SS14},
it is expected that the equivalence $\sigma_{3\mathrm{Q}}^\mathrm{Abel} 
\simeq \sigma_{3\mathrm{Q}}$ can be also realized in baryonic 3Q cases. 
To investigate the equivalence $\sigma_{3\mathrm{Q}}^\mathrm{Abel} \simeq \sigma_{3\mathrm{Q}}$,
it seems necessary to use 
(i) large numbers of gauge configurations, 
(ii) larger numbers of 3Q configurations, 
and (iii)  large-volume lattices of more than about $2$~fm, 
which are inspired from the analysis on the quark-confining force in mesons \cite{SS14}.
Therefore, in this paper, we perform the accurate calculation that meets 
the above conditions.
Then, we find equivalence 
$\sigma_{3\mathrm{Q}}^\mathrm{Abel} \simeq \sigma_{3\mathrm{Q}}$ 
within a few percent deviation.

\section{Numerical setting for maximal Abelian projection}
%
We perform the SU(3) quenched lattice QCD simulations with 
the standard plaquette action. 
We mainly use the lattice of $L^3 L_t=16^3 32$ at $\beta \equiv 6/g^2=5.8$, 
with the gauge coupling $g$, the spatial size $L^3$, 
and the temporal one $L_t$.
The lattice spacing is $a= 0.148(2)$~fm, which is determined 
so as to reproduce the string tension $\sigma =0.89$ GeV/fm 
in the Q$\bar{\mathrm{Q}}$ potential.
Thus, the physical spatial volume of the lattice is estimated as 
$(2.37(3)$~fm$)^3$. 
We also use a finer lattice of $20^3 32$ at $\beta =6.0$, 
which corresponds to $a=0.1022(5)$~fm and the physical spatial volume of 
$(2.05(1)$~fm$)^3$. 
The simulation conditions are summarized in Table~\ref{tab1}.

For $\beta$=5.8 and 6.0, 
we use $2000$ and $1000$ gauge configurations, 
respectively, which are taken every $500$ sweeps 
after a thermalization of $20000$ sweeps.
%
It is worth mentioning that the used configuration number $2000$ 
is about ten times larger than that 
in the previous detailed lattice studies of 
baryonic 3Q potentials \cite{TSNM01, TSNM02}.
The large number of the gauge configurations 
enables us to measure accurately the large-distance 3Q potential data, 
which is important for the confinement study.

In the lattice formalism, the SU(3) gauge field is described by
the link variable $U_\mu(s) =e^{iagA_\mu(s)} \in$ SU(3) 
instead of the gluon field $A_\mu(s) \in$ su(3).
We perform the SU(3) MA gauge fixing by maximizing
\begin{equation}
\begin{split}
 R_{\rm MA}[U_\mu(s)]
&\equiv \sum_{s} \sum_{\mu=1}^4  
{\rm tr}\left( U_\mu^\dagger(s)\vec H U_\mu(s)\vec H\right)  \\
&= \frac{1}{2} \sum_{s} \sum_{\mu=1}^4  
\left(  \sum_{i=1}^3 |U_\mu(s)_{ii} |^2 -1\right)
\end{split}
\label{MAgf}
\end{equation}
under the SU(3) gauge transformation 
$U_\mu(s) \mapsto \Omega(s)U_\mu(s)\Omega^\dagger(s+\hat\mu)$
with $\Omega(s) \in$ SU(3).
Here, $\vec H = (T_3, T_8)$ is the Cartan subalgebra of SU(3), 
and $T_3= {\rm diag}(1/2,-1/2,0)$ and 
$T_8=(1/2\sqrt{3}) \times {\rm diag}(1,1,-2)$ in the standard notation.
(The functional (\ref{MAgf}) has been used for MA gauge fixing 
in Refs.~\cite{DIK04_3Q, STW02, DIK04_QQ, BSW91, BD13, SS14}.)
We numerically maximize $R_{\rm MA}$ for each gauge configuration $\{U_\mu(s)\}$ until it converges,
by using the over-relaxation method \cite{STW02, SS14}. 
%
%
As for the stopping criterion, we stop the maximization algorithm, 
when the deviation $\Delta R_{\rm MA}/(4L^3L_t) < 10^{-9}$ after the one-sweep gauge transformation. 
From Eq.(\ref{MAgf}), we remark $-1/2 \leq R_{\rm MA}/(4L^3L_t) \leq 1$ for arbitrary gauge configuration $\{U_\mu(s)\}$.
The converged value of $\left< R_{\rm MA} \right>/(4L^3L_t)$ is 
$0.7072(6)$ at $\beta=5.8$ and $0.7322(5)$ at $\beta=6.0$, 
where $\langle \cdots \rangle$ is the statistical average over the gauge configurations and the value in parentheses denotes the standard deviation.
Note that the maximized value of $R_{\rm MA}$ is almost the same over 
1000--2000 gauge configurations because the standard deviation of $R_{\rm MA}$ is fairly small. 
Then, we expect that our procedure escapes bad local minima, 
where $R_{\rm MA}$ is relatively small, 
and the Gribov copy effect is not significant.
%
%
%

We extract the Abelian part of the link variable,
\begin{equation*}
u_\mu(s) = 
\exp \left( i\theta_\mu^3(s) T_3 + i\theta_\mu^8(s) T_8 \right)
\in {\rm U(1)}_3 \times {\rm U(1)}_8,
\end{equation*}
by maximizing the norm 
\begin{equation}
R_{\mathrm{Abel}} \equiv \frac{1}{3}
{\rm Re} \, {\rm tr}\left( U_\mu^{\rm MA}(s) u_\mu^\dagger(s) \right) 
\in [-\frac{1}{2},1],
\end{equation}
where $U_\mu^{\rm MA} (s) \in$ SU(3) denotes the link variable in the MA gauge.
In the MA gauge, there remains the residual ${\rm U(1)}^2$ gauge symmetry 
with the global Weyl (color permutation) symmetry \cite{IS99}.
In fact, $R_{\rm MA}$ in Eq.~(\ref{MAgf}) is invariant 
under the ${\rm U(1)}^2$ gauge transformation
$U_\mu(s)\mapsto \omega(s)U_\mu(s)\omega^\dagger(s+\hat\mu)$
with $\omega(s) \in$ U($1$)$_3 \times$U($1$)$_8$ 
and the global color permutation. 
Under the ${\rm U(1)}^2$ gauge transformation, the Abelian link variable $u_\mu(s)$ transforms as
\begin{equation}
u_\mu(s) \mapsto 
\omega(s)u_\mu(s)\omega^\dagger(s+\hat\mu),
\label{eq:U(1)2Gauge2}
\end{equation}
which means that  $u_\mu(s)$ behaves as a U(1)$^2$ gauge field.
Here, the MA-projected U(1)$^2$ Abelian theory is 
similar to compact QED, and it has not only the electric current 
but also the magnetic-monopole current.

Since off-diagonal-gluon components are suppressed in the MA gauge, 
we find approximate ``microscopic Abelian dominance'' \cite{IS99}
for the Abelian link variable as $u_\mu(s) \simeq U_\mu^{\rm MA}(s)$ or  
$\left<R_{\mathrm{Abel}}\right> \simeq 1$, i.e., 
$\left<R_{\mathrm{Abel}}\right> = 0.8924(3)$ at $\beta=5.8$ 
and $0.9027(2)$ at $\beta=6.0$.
However, it is a highly nontrivial question 
whether this gauge shows ``macroscopic Abelian dominance'' such as 
Abelian dominance of quark confinement in Q$\bar{\mathrm{Q}}$ and 
3Q potentials.

\begin{table}[t]
\caption{The simulation condition: $\beta$, the lattice size $L^3L_t$, 
and the gauge-configuration number $N_{\rm con}$. 
The corresponding lattice spacing $a$ and the physical spatial size 
$La$ are also listed. 
Here, the values in parentheses denote the standard error.
}
\label{tab1}
\begin{ruledtabular}
\begin{tabular}{cccll}
$\beta$ & $L^3 L_t$ & $N_{\rm con}$ & \, $a$ [fm] &  $La$ [fm]  \\
\hline
  $5.8$   & $16^332$     & 2000    & 0.148(2)   & 2.37(3)    \\
  $6.0$   & $20^332$     & 1000    & 0.1022(5)  & 2.05(1)    \\
\end{tabular} 
\end{ruledtabular}
 \end{table}

\section{Numerical calculation method for three-quark potential}
Similar to the case of the Q$\bar{\mathrm{Q}}$ potential $V(r)$ \cite{SS14}, 
the color-singlet baryonic 3Q potential $V_{3\mathrm{Q}}$ 
can be calculated as \cite{TSNM01,TSNM02,VVG14} 
\begin{equation}
V_{3\mathrm{Q}} = - \lim_{T\rightarrow \infty}\frac{1}{T}\ln \left\langle 
W_{3\mathrm{Q}} \left[ U_\mu(s)  \right] \right\rangle
\end{equation}
from the 3Q Wilson loop 
\begin{equation}
W_{3Q} \left[ U_\mu(s)  \right]
\equiv \frac{1}{3!} \sum_{a,b,c} \sum_{a^\prime b^\prime c^\prime} \epsilon_{abc}\epsilon_{a^\prime b^\prime c^\prime}
X_1^{aa^\prime}X_2^{bb^\prime}X_3^{cc^\prime}.
\end{equation}
Here, $X_k \equiv \prod_{\Gamma_k}  U_\mu (s)$
is the path-ordered product of the link variables along the path denoted by $\Gamma_k$ in Fig.~\ref{Fig:1}(b). 
The 3Q Wilson loop represents that the gauge-invariant 3Q state is generated at $t=0$ and is annihilated at $t=T$ with the three quarks spatially fixed in $\mathbb{R}^3$ for $0<t<T$. 
We note that the potential $V_{3\mathrm{Q}}$ is independent of 
the choice of the junction point $O$ \cite{TSNM01, TSNM02}, 
which is different from the physical junction at the Fermat point.

As shown in Fig.~\ref{Fig:1}(c), we put three quarks on $(i,0,0)$, $(0,j,0)$, and $(0,0,k)$ in $\mathbb{R}^3$ with $1\leq i \leq j \leq k \leq L/2$ in lattice units
and set the junction point $O$ at the origin $(0,0,0)$.
For the calculation of the 3Q Wilson loop, 
we use the translational, the rotational, and the reflection symmetries on the lattices.
%
Here, we deal with 101 and 211 different patterns of 3Q systems at $\beta$=5.8 and 6.0, respectively,
based on well-converged data of $\left\langle W_{3\mathrm{Q}} \right\rangle$. 

We extract $V_{3\mathrm{Q}}$ from the least-squares fit with the single-exponential form
$\left< W_{3\mathrm{Q}}(T) \right> = \tilde{C} e^{-V_{3\mathrm{Q}} T}$.
Here, we choose the fit range of $T_\textrm{min} \leq T \leq T_\textrm{max}$ such that the stability of the so-called effective mass
\begin{equation}
V_{3\mathrm{Q}}^\textrm{eff}(T) \equiv \ln 
\frac{\left< W_{3\mathrm{Q}}(T) \right>}{\left< W_{3\mathrm{Q}}(T+1) \right>}
\label{sup-eq:eff-mass}
\end{equation}
is observed in the range $T_\textrm{min} \leq T \leq T_\textrm{max}-1$.
On the error estimate, we use the jackknife method.
%
%

For the accurate calculation of the 3Q potential with finite $T$, we apply here the gauge-invariant smearing method \cite{APE87,BSS93,TSNM01, TSNM02}, 
which enhances the ground-state component in the 3Q state in $W_{3\mathrm{Q}}$.
The smearing is performed as the iterative replacement 
of the spatial link variables $U_i(s)$ $(i \in 1,2,3)$ 
by the obscured link variables $\bar{U}_i(s) \in$ SU($3$) 
which maximizes 
$\mathrm{Re} \, \mathrm{tr} \left[ \bar{U}_i^\dagger(s) V_i(s) \right]$ with
\begin{equation}
V_i(s) \equiv \alpha U_i(s) + \sum_{j\not = i} \sum_{\pm} U_{\pm j}(s) U_i(s\pm \hat{j})U_{\pm j}^\dagger(s\pm \hat{i}).
\end{equation}
Here, we denote $U_{-j}(s) \equiv U_j^\dagger(s-\hat{j})$.
(For the details of the smearing method, 
see Secs.~III.B and III.C in Ref.~\cite{TSNM02}.)
%
For the case of $\beta=5.8$ on $16^332$, we adopt the smearing parameter $\alpha =2.3$ and choose the iteration number $N_\textrm{smr}=25$ and $4$ for SU(3) QCD and the Abelian part, respectively,
so as to largely enhance the ground-state overlap for each part.
We have confirmed that the results are almost unchanged 
by changing the iteration number $N_{\rm smr}$.

\begin{figure}[t]
\centering
\includegraphics[width=8.6cm,clip]{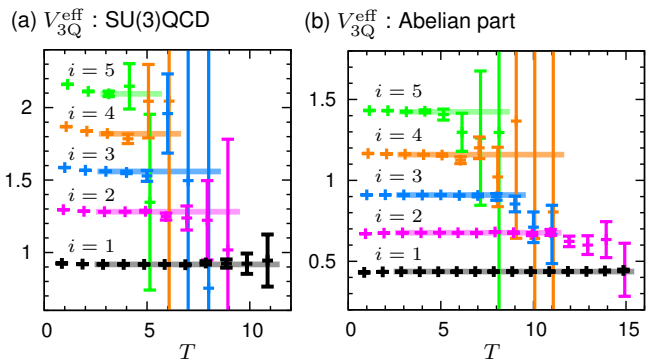}
\caption{Typical examples of effective mass plots of the 3Q potential 
for (a) SU(3) QCD and (b) the Abelian part in lattice units. 
Here, three quarks are put on the equilateral-triangle configuration, i.e., 
$i=j=k=1, \dots, 5$ in Fig.~{\ref{Fig:1}}(c), 
on a $16^332$ lattice at $\beta$=5.8.
The solid horizontal lines denote the obtained values of $V_{\mathrm{3Q}}$ and $V_{\mathrm{3Q}}^{\mathrm{Abel}}$
and are extended in the corresponding fit range of $T_\textrm{min} \leq T \leq T_\textrm{max}-1$.
}
 \label{Fig:Veff3Q}
\end{figure}

Similarly, we also calculate the MA projection of the 3Q potential
\begin{eqnarray}
V_{3\mathrm{Q}}^{\mathrm{Abel}} = - \lim_{T\rightarrow \infty}
\frac{1}{T}\ln \left\langle 
W_{3\mathrm{Q}} \left[ u_\mu(s)  \right] \right\rangle
\end{eqnarray}
from the Abelian 3Q Wilson loop in the MA gauge, 
$W_{3Q} \left[ u_\mu(s)\right]$, which is invariant 
under the residual Abelian gauge transformation~(\ref{eq:U(1)2Gauge2}).
By way of illustration, we show in Fig.~\ref{Fig:Veff3Q} 
the effective mass plot for each part at $\beta$=5.8 on $16^332$.

\begin{figure*}[t]
\centering
\includegraphics[width=17.8cm,clip]{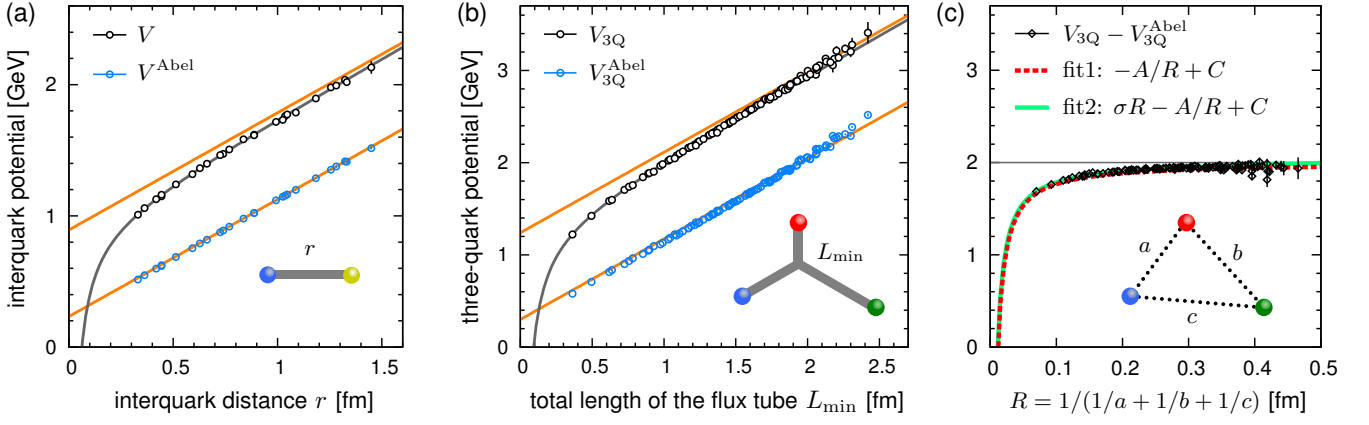}
\caption{
MA projection of (a) ${\rm Q}\bar{\rm Q}$ and (b) 3Q potentials 
in SU(3) quenched lattice QCD at $\beta$=5.8 on $16^332$. 
In each panel, the black and blue circles denote the original SU($3$) potential and the Abelian part, respectively.
The gray curves are obtained by the best fit with Eqs.~(\ref{eq:Cornell}) and (\ref{eq:Y-ansatz_ET}) (Y ansatz), as listed in Table~\ref{tab:fitmain}.
The slopes of the \textit{parallel} orange 
lines for the panels~(a) and (b) are 
$\sigma$ and $\sigma_{3\mathrm{Q}}$, respectively.
(c) Fit analysis of $V_{3\mathrm{Q}} - V_{3\mathrm{Q}}^{\mathrm{Abel}}$ 
(black diamonds) to illustrate the equivalence 
$\sigma_{\mathrm{3Q}} \simeq \sigma_{\mathrm{3Q}}^{\mathrm{Abel}}$ 
between the three-body string tension and its Abelian part
in the baryonic 3Q potential.
The red dashed and the green solid curves are the best fit with the pure Coulomb ansatz (\ref{eq:Coulomb-ansatz-DV}) and the Coulomb-plus-linear ansatz, respectively.
}
\label{fig:main}
\end{figure*}


\begin{table*}[t]
\caption{Fit analysis of interquark potentials 
in lattice units 
at $\beta =5.8$ (i.e., $a\simeq 0.15$~fm) on $16^3 32$ 
and $\beta=6.0$ (i.e., $a\simeq 0.10$~fm) on $20^3 32$.
The best-fit parameter sets $(\sigma, A, C)$ of the Q$\bar{\rm Q}$ potential 
$V$ and the Abelian part $V^{\mathrm{Abel}}$ are listed 
with the functional form~(\ref{eq:Cornell}).
The best-fit parameter sets 
$(\sigma_{3\mathrm{Q}}, A_{3\mathrm{Q}}, C_{3\mathrm{Q}})$ 
of the 3Q potential $V_{3\mathrm{Q}}$ and 
the Abelian part $V_{3\mathrm{Q}}^{\mathrm{Abel}}$ 
are listed with the Y ansatz (\ref{eq:Y-ansatz}).
The label of (equi.~triangle) means the fit analysis only with 
the lattice data of equilateral-triangle 3Q configurations.
$N_Q$ is the number of different patterns of Q$\bar{\mathrm{Q}}$ or 3Q systems.
The string tension ratio $\sigma^\mathrm{Abel}/\sigma$ is listed at the last column.
%
\label{tab:fitmain}}
\begin{ruledtabular}
\begin{tabular}{clc lll lll c}
 & &
 & \multicolumn{3}{c}{SU(3)} 
 & \multicolumn{3}{c}{Abelian part}  \\
\cline{4-6}\cline{7-9}
$\beta$ & 
 & $N_\mathrm{Q}$
 &  \qquad $\sigma $  & \,\,$A$ & \,\,$C$ 
 &  \quad $\sigma^\mathrm{Abel} $  & $A^\mathrm{Abel}$ & $C^\mathrm{Abel}$ 
 &  $\sigma^\mathrm{Abel}/\sigma $    \\
\hline 
5.8 
    & Q$\bar{\mathrm{Q}}$ & 26 
    & \,\, 0.099(2) & 0.30(3)   & 0.67(2)
    & \,\, 0.098(1) & 0.043(12)   & 0.187(7)  
    & \,\, 0.99(3)       \\
& 3Q (equi.~triangle) & 5
    & \,\, 0.097(1) & 0.118(3) & 0.93(1) 
    & \,\, 0.098(3) & $-$0.001(8) & 0.19(2)
    & \,\, 1.01(3)       \\
& 3Q & 101
    & \,\, 0.0997(4) & 0.109(1) & 0.905(4) 
    & \,\, 0.0967(5) & 0.006(2)  &  0.213(5)
    & \,\, 0.97(1)       \\ 
\\
6.0 
    & Q$\bar{\mathrm{Q}}$ & 39 
    & \,\, 0.0472(6) & 0.289(10)   & 0.658(5)
    & \,\, 0.0457(2) & 0.050(3)   & 0.183(2)  
    & \,\, 0.97(1)       \\
& 3Q (equi.~triangle) & 8
    & \,\, 0.0471(10) & 0.121(3) & 0.936(9) 
    & \,\, 0.0455(12) & 0.014(4)  &  0.233(12)
    & \,\, 0.97(3)       \\
& 3Q & 211
    & \,\, 0.0480(3) & 0.113(1) & 0.917(3) 
    & \,\, 0.0456(2) & 0.013(1) & 0.232(2)
    & \,\, 0.95(1)       \\
\end{tabular}
\end{ruledtabular}
\end{table*}




\section{Abelian dominance of quark confinement in 3Q potential}
%
In this section, we show the numerical results of 
Q$\bar{\mathrm{Q}}$ and 3Q systems 
in SU(3) quenched lattice QCD at $\beta$=5.8 on $16^332$.
Figure~\ref{fig:main}(a) shows the Q$\bar{\mathrm{Q}}$ potential $V(r)$ 
and the Abelian part $V^{\mathrm{Abel}}(r)$.
All the lattice data of $V(r)$ are well reproduced by the Coulomb-plus-linear ansatz~(\ref{eq:Cornell}) 
with the best-fit parameter set listed in Table~\ref{tab:fitmain}.
For a larger interquark distance $r$ than $1$~fm, $V(r)$ is simply described 
by the linear quark-confining potential $\sigma r +C$ 
[upper straight line in Fig.~\ref{fig:main}(a)].
Figure~\ref{fig:main}(a) illuminates ``perfect 
Abelian dominance'' of confinement in the Q$\bar{\mathrm{Q}}$ potential, 
which was reported in Ref.~\cite{SS14}, because the Abelian part 
$V^{\mathrm{Abel}}(r)$ has a significant agreement with $\sigma r +C'$ 
[lower straight line in Fig.~\ref{fig:main}(a)] at large distances.


We note that the Abelian dominance of the Q$\bar{\mathrm{Q}}$-confining force 
does not necessarily mean that of the 3Q-confining force
because one cannot superpose solutions in QCD even at the classical level. 
Indeed, a 3Q system cannot be described by the superposition of 
the interaction between two quarks, 
as is suggested from the functional form (\ref{eq:Y-ansatz})
of the 3Q potential \cite{TSNM01, TSNM02}.
We find, however, Abelian dominance 
of the 3Q-confining force with an accuracy within a few percent deviation as described below.

Figure~\ref{fig:main}(b) shows the 3Q potential 
$V_{3\mathrm{Q}}$ and the Abelian part $V_{3\mathrm{Q}}^{\mathrm{Abel}}$ 
plotted against the total length of the flux tube, $L_{\mathrm{min}}$.
All the lattice data of $V_{3\mathrm{Q}}$ are approximately described by 
a single-valued function of $L_{\mathrm{min}}$, 
although $V_{3\mathrm{Q}}$ generally depends on 
the relative position of the three quarks. 
The main reason is that the three-body confinement term 
$\sigma_{3\mathrm{Q}} L_{\mathrm{min}}$  
is relevant in the Y ansatz~(\ref{eq:Y-ansatz}) except for short distances. 
When the 3Q system forms an equilateral triangle, one finds 
$L_{\mathrm{min}}=\sqrt{3}\left|\mathbf{r}_i - \mathbf{r}_j\right|$ 
for any $i\ne j$,
and the Y ansatz (\ref{eq:Y-ansatz}) becomes
\begin{equation}
V_{3\mathrm{Q}} ({\bf r}_1, {\bf r}_2, {\bf r}_3)
= \sigma_{3\mathrm{Q}}  L_{\rm min}- 
3\sqrt{3} \, \frac{ A_{3\mathrm{Q}}}{L_{\mathrm{min}}}
+C_{3\mathrm{Q}}.
\label{eq:Y-ansatz_ET}
\end{equation}
Since $V_{3\mathrm{Q}}$ approximately obeys 
a single-valued function of $L_{\rm min}$, 
all the lattice data are well reproduced by Eq.~(\ref{eq:Y-ansatz_ET}) with the best-fit parameter set as listed in Table~\ref{tab:fitmain}, other than the equilateral-triangle 3Q systems.
 %
 %
When the total flux-tube length $L_{\mathrm{min}}$ is larger than $1$~fm, 
$V_{3\mathrm{Q}}$ is described by the linear 3Q-confining potential $\sigma_{3\mathrm{Q}} L_{\mathrm{min}} +C_{3\mathrm{Q}}$ 
[upper straight line in Fig.~\ref{fig:main}(b)].
Remarkably, the Abelian part $V^{\mathrm{Abel}}(r)$ has a significant agreement with $\sigma_{3\mathrm{Q}} L_{\mathrm{min}} +C_{3\mathrm{Q}}'$ 
[lower straight line in Fig.~\ref{fig:main}(b)] at large distances,
which is plausible evidence for $\sigma_{\rm 3Q}^{\rm Abel} \simeq \sigma_{\rm 3Q}$ in the baryonic 3Q potential.

To demonstrate 
$\sigma_{\rm 3Q}^{\rm Abel} \simeq \sigma_{\rm 3Q}$ conclusively, we investigate the difference between
$V_{3\mathrm{Q}}$ and $V_{3\mathrm{Q}}^{\mathrm{Abel}}$ at long distances as shown in Fig.~\ref{fig:main}(c).
As is the case in $V_{3\mathrm{Q}}$, the Abelian part of the 3Q potential 
has the functional form 
\begin{eqnarray}
V_{3\mathrm{Q}}^{\mathrm{Abel}}
= \sigma_{3\mathrm{Q}}^{\mathrm{Abel}}  L_{\rm min}
- \frac{A_{3\mathrm{Q}}^{\mathrm{Abel}}}{R}
+C_{3\mathrm{Q}}^{\mathrm{Abel}},
\end{eqnarray}
where $1/R \equiv \sum_{i<j} 1/|{\bf r}_i-{\bf r}_j|$ \cite{BKV13}.
If the Abelian dominance of the 3Q potential is exact, i.e., 
$\sigma_{3\mathrm{Q}}^{\mathrm{Abel}} = \sigma_{3\mathrm{Q}}$, 
one has to observe 
\begin{equation}
\Delta V_{3\mathrm{Q}} \equiv V_{3\mathrm{Q}} - V_{3\mathrm{Q}}^{\mathrm{Abel}}
= - \frac{\Delta A_{3\mathrm{Q}}}{R} + \Delta C_{3\mathrm{Q}},
\label{eq:Coulomb-ansatz-DV}
\end{equation}
where $\Delta A_{3\mathrm{Q}}\equiv A_{3\mathrm{Q}}- A_{3\mathrm{Q}}^{\mathrm{Abel}}$
and $\Delta C_{3\mathrm{Q}}\equiv C_{3\mathrm{Q}}-C_{3\mathrm{Q}}^{\mathrm{Abel}}$.
Then, we try a fit analysis to 
$\Delta V_{3\mathrm{Q}}$ with the pure Coulomb ansatz~(\ref{eq:Coulomb-ansatz-DV}) (fit 1) 
and the Coulomb-plus-linear ansatz, $\Delta \sigma_{3\mathrm{Q}}' R - \Delta A_{3\mathrm{Q}}'/R + \Delta C_{3\mathrm{Q}}'$ (fit 2), in Fig.~\ref{fig:main}(c).
Fits 1 and 2 reveal that $\Delta V_{3\mathrm{Q}}$ has almost zero string tension, $\Delta \sigma_{3\mathrm{Q}}' \simeq 0$, and is well reproduced by the pure Coulomb ansatz (\ref{eq:Coulomb-ansatz-DV}).
Therefore, we conclude that there is no difference between the string tensions in 
$V_{3\mathrm{Q}}$ and $V_{3\mathrm{Q}}^{\mathrm{Abel}}$, 
i.e., $\sigma_{\rm 3Q}^{\rm Abel} \simeq \sigma_{\rm 3Q}$, 
with an accuracy within a few percent deviation.
[We remark that Fig.~\ref{fig:main}(c) would be plausible evidence for the exact equivalence of $\sigma_{\rm 3Q}^{\rm Abel} = \sigma_{\rm 3Q}$.]

\begin{figure}[t]
\centering
\includegraphics[width=8.5cm,clip]{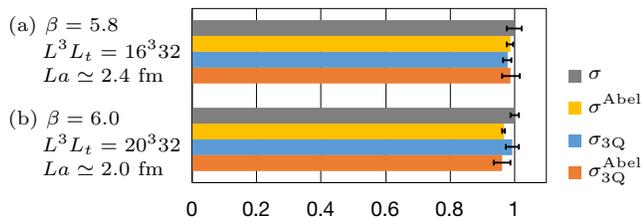}
\caption{
Comparison of string tensions of the Q$\bar{\mathrm{Q}}$ and 3Q potentials for SU(3) QCD and the Abelian part.
Here, $\sigma$ and $\sigma_{\rm Abel}$ are the string tensions of the Q$\bar{\mathrm{Q}}$ potential for SU(3) QCD and the Abelian part, respectively.
Similarly, $\sigma_{\rm 3Q}$ and $\sigma_{\rm 3Q}$ are the string tensions of the 3Q potential for SU(3) QCD and the Abelian part, respectively.
In both cases of (a) $\beta =5.8$  on the $16^332$ lattice and (b) $\beta=6.0$ on the $16^332$ lattice, we find that the string tensions are equivalent within a few percent deviation:
$\sigma \simeq \sigma^{\rm Abel} \simeq \sigma_{\rm 3Q} \simeq \sigma_{\rm 3Q}^{\rm Abel}$.
%
%
 }
 \label{Fig:Summary}
\end{figure}

\begin{figure}[t]
\includegraphics{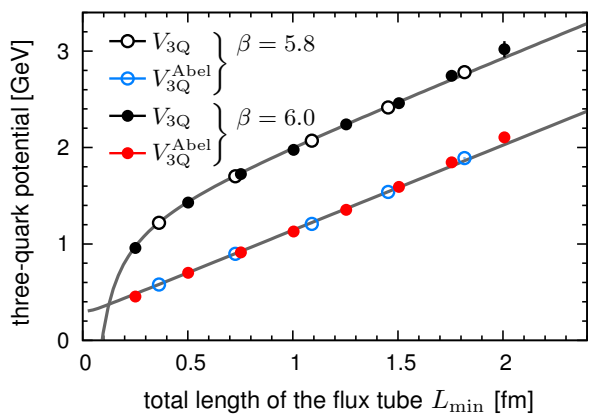}
\caption{
The MA projection of the 3Q potential 
for equilateral-triangular configurations 
plotted against $L_{\rm min}$ 
for $\beta=5.8$ and $6.0$ in the physical unit. 
The curves are obtained by the best fit  
with Eq.~(\ref{eq:Y-ansatz_ET}) for the $\beta=5.8$ data, 
as listed in Table~\ref{tab:fitmain}.
}
\label{fig:60}
\end{figure}

%

To see the finite lattice-spacing effect, 
we also perform SU(3) quenched lattice QCD at $\beta=6.0$ on $20^3 32$ and summarize the results in Table~\ref{tab:fitmain} and Fig.~\ref{Fig:Summary}(b).
We find again $\sigma_{\rm 3Q}^{\rm Abel} \simeq \sigma_{\rm 3Q}$, and thus 
$\sigma_{\rm 3Q}^{\rm Abel} \simeq \sigma_{\rm 3Q}$ is expected to be true in the 
finer lattice spacings, similar to the case of the Q$\bar{\textrm{Q}}$ potential \cite{SS14}.
Figure~\ref{Fig:Summary} compares the Q$\bar{\textrm{Q}}$ and 3Q results at $\beta=5.8$ on the $16^332$ lattice and  at $\beta=6.0$ on the $20^332$ lattice.
In both cases, we find that the string tensions of the Q$\bar{\textrm{Q}}$ and 3Q potentials for the SU(3) and Abelian part are equivalent within a few percent deviation:
$\sigma \simeq \sigma^{\rm Abel} \simeq \sigma_{\rm 3Q} \simeq \sigma_{\rm 3Q}^{\rm Abel}$.
To be exact, in the case of $\beta=6.0$ on the $20^332$ lattice, 
the string tensions of the Abelian part ($\sigma^{\rm Abel} \simeq \sigma_{\rm 3Q}^{\rm Abel}$) are about $3 \%$ smaller than SU(3) QCD ($\sigma \simeq \sigma_{\rm 3Q}$) because the physical spatial size is slightly small ($La\simeq 2.0$~fm).
This physical-spatial-size effect is expected from the result of the Q$\bar{\textrm{Q}}$ potential \cite{SS14} (see Fig.~\ref{Fig:PRDR}).

For a visual demonstration, 
we restrict ourselves on equilateral-triangular 3Q configurations
and show their lattice results of $V_{\rm 3Q}$ and $V_{\rm 3Q}^{\rm Abel}$ 
at $\beta$=5.8 and 6.0 in the physical unit in Fig.~\ref{fig:60}, 
where an irrelevant constant is shifted.
For each of $V_{\rm 3Q}$ and $V_{\rm 3Q}^{\rm Abel}$, 
both lattice data are found to be well reproduced by a single curve.
We list their raw data in Table~\ref{tab2} and add the fit result in Table~\ref{tab:fitmain}.

To conclude, we thus find Abelian dominance of the string tension, 
$\sigma_{\rm 3Q}^{\rm Abel} \simeq \sigma_{\rm 3Q}$, with an accuracy within a few percent deviation
in the baryonic 3Q potential in SU(3) quenched lattice QCD 
for $\beta$=5.8 on $16^3 32$ and $\beta$=6.0 
on $20^3 32$, as shown in Table~\ref{tab:fitmain}.

\begin{table}[t]
\caption{A part of the lattice data of the 3Q potential $V_{\rm 3Q}$ 
and the Abelian part $V_{\rm 3Q}^{\rm Abel}$ restricted for 
the equilateral-triangle configuration, 
i.e., $i=j=k$ in Fig.~{\ref{Fig:1}}(c), in lattice units.
}
\label{tab2}
\begin{ruledtabular}
\begin{tabular}{cr llll}
 & & \multicolumn{2}{c}{$16^332$ at $\beta=5.8$} 
 & \multicolumn{2}{c}{$20^332$ at $\beta=6.0$}  \\
\cline{3-4}\cline{5-6}
 $(i,j,k)$ & $L_\mathrm{min}$ & $V_{3\mathrm{Q}}$
 & $V_{3\mathrm{Q}}^\mathrm{Abel}$ & $V_{3\mathrm{Q}}$ & $V_{3\mathrm{Q}}^\mathrm{Abel}$ \\
\hline
$(1,1,1)$ &   2.45 &  0.9176(2) &  0.4361(1) & 0.7943(3)& 0.3140(1) \\
$(2,2,2)$ &   4.90 &  1.2812(9) &  0.6765(4) & 1.0393(8)& 0.4425(2)  \\
$(3,3,3)$ &   7.35 &  1.559(2)  &   0.9095(9) & 1.193(2)  & 0.5521(3)  \\
$(4,4,4)$ &   9.80 &  1.819(6)  &   1.159(2)    & 1.323(3)&  0.6648(6)  \\
$(5,5,5)$ & 12.24 &  2.10(2)    &   1.424(4)    & 1.460(6)&  0.782(1)  \\
$(6,6,6)$ & 14.70 & $\quad \cdots$ & $\quad \cdots$ & 1.58(1)& 0.906(2)  \\
$(7,7,7)$ & 17.15 & $\quad \cdots$ & $\quad \cdots$ & 1.72(1)&   1.037(2) \\
$(8,8,8)$ & 19.60 & $\quad \cdots$ & $\quad \cdots$ & 1.87(4)&   1.172(3)\\
\end{tabular} 
\end{ruledtabular}
 \end{table}

\section{Summary and concluding remarks}

We have studied the MA projection of quark confinement 
in the baryonic 3Q potential in the SU(3) quenched lattice QCD 
with $\beta=5.8$ on $16^3 32$ and $\beta=6.0$ on $20^3 32$ 
for more than 300 different 3Q systems in total, 
with 1000--2000 gauge configurations.
(Note also that the lattice data of $V_{\rm 3Q}$ themselves 
are fairly accurate, because of the high statistics.)
Remarkably, we have found Abelian dominance of 
the string tension with an accuracy within 
a few percent deviation, $\sigma \simeq \sigma^{\rm Abel} \simeq \sigma_{\rm 3Q} \simeq \sigma_{\rm 3Q}^{\rm Abel}$, in Q$\bar{\mathrm{Q}}$ and 3Q potentials simultaneously 
on these lattices.
(For a more definite conclusion, it is desired to perform 
similar studies with larger and finer lattices.)
Thus, despite the non-Abelian nature of QCD, quark confinement is 
entirely kept in the Abelian sector of QCD in the MA gauge.
In other words, Abelianization of QCD can be realized 
without the loss of the quark-confining force via the MA projection.
This fact would be meaningful to understand 
the confinement mechanism in the non-Abelian gauge theory of QCD.
Furthermore, the Abelian dominance for both Q$\bar{\mathrm{Q}}$ 
and 3Q potentials indicates a universality of the confinement mechanism 
for the wide category of hadrons in terms of Abelianization of QCD.

\begin{acknowledgments}
The authors thank Hideaki~Iida and Toru~T.~Takahashi. 
H.~S. thanks V.~G.~Bornyakov for his valuable suggestions.
N.~S. is supported by a Grant-in-Aid for JSPS Fellows and by JSPS KAKENHI Grant 250588, 15K17725.
H.~S. is supported by the Grant for Scientific Research 
[(C) Grant No.23540306 and No.15K05076] from the Ministry of Education, 
Science and Technology of Japan.
The lattice calculations were partially performed on NEC-SX8R at Osaka University.
This work was partially supported by RIKEN iTHES Project.
\end{acknowledgments}


\begin{thebibliography}{50}
\bibitem{QCD} 
H.~J.~Rothe, {\it Lattice Gauge Theories}, 4th ed. (World Scientific, Singapore, 2012), and references therein.

\bibitem{TSNM01} 
T.~T.~Takahashi, H.~Matsufuru, Y.~Nemoto, and H.~Suganuma, Phys. Rev. Lett. {\bf 86}, 18 (2001); 
T.~T.~Takahashi and H.~Suganuma, Phys. Rev. Lett. {\bf 90}, 182001 (2003); Phys. Rev. D {\bf 70}, 074506 (2004).

\bibitem{TSNM02} T.~T.~Takahashi, H.~Suganuma, Y.~Nemoto, and H.~Matsufuru, Phys. Rev. D {\bf 65}, 114509 (2002).

\bibitem{C2004}
J.~M.~Cornwall, Phys. Rev. D {\bf 69}, 065013 (2004).

\bibitem{BSS93} G.~S.~Bali and K.~Schilling, Phys. Rev. D {\bf 47}, 661 (1993).

\bibitem{B01} For a review article, see G.~S.~Bali, Phys. Rep. {\bf 343}, 1 (2001).

\bibitem{OST05} F.~Okiharu, H.~Suganuma, and T.~T.~Takahashi, Phys. Rev. D {\bf 72}, 014505 (2005).

 

\bibitem{CNN79} A.~Casher, H.~Neuberger, and S.~Nussinov, Phys. Rev. D {\bf 20}, 179 (1979). 

\bibitem{IBSS03} H.~Ichie, V.~Bornyakov, T.~Streuer, and G.~Schierholz, Nucl. Phys. {\bf A721}, C899 (2003).
\bibitem{DIK04_3Q} V.~G.~Bornyakov {\it et al.} (DIK Collaboration), Phys. Rev. D {\bf 70}, 054506 (2004).
\bibitem{BS04} P.~O.~Bowman and A.~P.~Szczepaniak, Phys. Rev. D {\bf 70}, 016002 (2004).
\bibitem{CCB13} 
M.~Cardoso, N.~Cardoso, and P.~Bicudo, Phys. Rev. D {\bf 86}, 014503 (2012);
N.~Cardoso, M.~Cardoso, and P.~Bicudo, Phys. Rev. D {\bf 88}, 054504 (2013); 

\bibitem{CCCP14} 
M.~S.~Cardaci, P.~Cea, L.~Cosmai, R.~Falcone, and A.~Papa, Phys. Rev. D {\bf 83}, 014502 (2011).


\bibitem{A08} O.~Andreev, Phys. Rev. D {\bf 78}, 065007 (2008). 

\bibitem{NtHM} Y.~Nambu, Phys. Rev. D {\bf 10}, 4262 (1974); 
G.~'t~Hooft, in {\it High Energy Physics}, 
(Editorice Compositori, Bologna, 1975); 
S.~Mandelstam, Phys. Rep. {\bf 23}, 245 (1976).

\bibitem{SW94} N.~Seiberg and E.~Witten, 
Nucl. Phys. {\bf B426}, 19 (1994); {\bf B431}, 484 (1994).

\bibitem{tH81} G.~'t~Hooft, Nucl. Phys. {\bf B190}, 455 (1981).

\bibitem{EI82} Z.~F.~Ezawa and A.~Iwazaki, Phys. Rev. D {\bf 25}, 2681 (1982). 

\bibitem{KSW87} 
A.~S.~Kronfeld, G.~Schierholz, and U.-J.~Wiese, Nucl. Phys. {\bf B293} 461 (1987); 
A.~S.~Kronfeld, M.~L.~Laursen, G.~Schierholz, and U.-J.~Wiese, Phys. Lett. B {\bf 198}, 516 (1987).

\bibitem{SY90} T.~Suzuki and I.~Yotsuyanagi, Phys. Rev. D {\bf 42}, 4257 (1990).

\bibitem{SNW94} J.~D.~Stack, S.~D.~Neiman, and R.~J.~Wensley, Phys. Rev. D {\bf 50}, 3399 (1994).

\bibitem{AS99} K.~Amemiya and H.~Suganuma, Phys. Rev. D {\bf 60}, 114509 (1999).

\bibitem{M95} O. Miyamura, Phys. Lett. B {\bf 353}, 91 (1995).

\bibitem{W95} R.~M.~Woloshyn, Phys. Rev. D {\bf 51}, 6411 (1995).

\bibitem{S95} H. Suganuma, A. Tanaka, S. Sasaki, and O. Miyamura,
Nucl. Phys. {\bf B} (Proc. Suppl.) {\bf 47}, 302 (1996).

\bibitem{STW02} J.~D.~Stack, W.~W.~Tucker, and R.~J.~Wensley, Nucl. Phys. {\bf B639}, 203 (2002).
\bibitem{DIK04_QQ} V.~G.~Bornyakov {\it et al.} (DIK Collaboration), Phys. Rev. D {\bf 70}, 074511 (2004).


\bibitem{SS14} N.~Sakumichi and H.~Suganuma, Phys. Rev. D {\bf 90}, 111501 (2014).

\bibitem{KKS15} 
S.~Kato, K.-I.~Kondo, and A.~Shibata, Phys. Rev. D {\bf 91}, 034506 (2015).


\bibitem{BSW91} F.~Brandstater, G.~Schierholz, U.-J.~Wiese, Phys. Lett. B {\bf 272}, 319 (1991). 
\bibitem{BD13} C.~Bonati and M.~D'Elia, Nucl. Phys. {\bf B877}, 233 (2013).


\bibitem{IS99} H.~Ichie and H.~Suganuma, Nucl. Phys. {\bf B548}, 365 (1999);
Phys. Rev. D {\bf 60}, 077501 (1999).

\bibitem{VVG14} J.~Vijande, A.~Valcarce, and H.~Garcilazo, Phys. Rev. D {\bf 90}, 094004 (2014).

\bibitem{APE87} M.~Albanese {\it et al.} (APE Collaboration), Phys. Lett. B {\bf 192}, 163 (1987). 

\bibitem{BKV13} N.~Brambilla, J.~Ghiglieri, and A.~Vairo, Phys. Rev. D {\bf 81}, 054031 (2010); 
N.~Brambilla, F.~Karbstein, and A.~Vairo, Phys. Rev. D {\bf 87}, 074014 (2013).












\end{thebibliography}
\end{document}